\documentclass{article}

\usepackage{arxiv}

\usepackage[utf8]{inputenc} 
\usepackage[T1]{fontenc}    
\usepackage{hyperref}       
\usepackage{url}            
\usepackage{booktabs}       
\usepackage{amsfonts}       
\usepackage{nicefrac}       
\usepackage{microtype}      
\usepackage{lipsum}		
\usepackage{graphicx}
\usepackage{upgreek}
\usepackage{soul}
\usepackage{xcolor}
\usepackage{natbib}
\usepackage{doi}
\usepackage{amsmath}
\usepackage{amssymb}

\title{Non-destructive depth reconstruction of Al-Al\textsubscript{2}Cu layer structure with nanometer resolution using extreme ultraviolet coherence tomography}

\usepackage{authblk}

\title{Non-destructive depth reconstruction of Al-Al\textsubscript{2}Cu layer structure with nanometer resolution using extreme ultraviolet coherence tomography}
\author[1,*]{Johann J. Abel}
\author[2,3,*]{Jonathan Apell }
\author[1]{Felix Wiesner}
\author[1,4]{Julius Reinhard}
\author[1,4,5]{Martin W\"unsche}
\author[6]{Nadja Felde}
\author[7]{Gabriele Schmidl}
\author[7]{Jonathan Plentz}
\author[1,4]{Gerhard G. Paulus}
\author[2]{Stephanie Lippmann}
\author[1,4,5,8]{Silvio Fuchs}

\affil[1]{Institute of Optics and Quantum Electronics, Friedrich Schiller University Jena, Max-Wien-Platz 1, 07743 Jena, Germany}
\affil[2]{Otto Schott Institute of Materials Research, Friedrich Schiller University Jena, Löbdergraben 32, 07743 Jena, Germany}
\affil[3]{Institute of Materials Science and Engineering, Chemnitz University of Technology, Erfenschlager Str. 73, 09125 Chemnitz, Germany}
\affil[4]{Helmholtz Institute Jena, GSI Helmholtzzentrum für Schwerionenforschung GmbH, Fraunhofer Str. 8, 07743 Jena, Germany}
\affil[5]{Indigo Optical Systems GmbH, Moritz-von-Rohr-Str. 1a,
07745 Jena, Germany}
\affil[6]{Fraunhofer Institute for Applied Optics and Precision Engineering IOF, Albert-Einstein-Straße 7, 07745 Jena, Germany}
\affil[7]{Leibniz Institute of Photonic Technology e.V., Albert-Einstein Straße 9, 07745 Jena, Germany}
\affil[8]{Laserinstitut Hochschule Mittweida, University of Applied Science Mittweida, Technikumplatz 17, 09648 Mittweida, Germany}

\renewcommand{\headeright}{}
\renewcommand{\undertitle}{}
\renewcommand{\shorttitle}{Non-destructive depth reconstruction of Al-Al\textsubscript{2}Cu layer structure with nanometer resolution using extreme ultraviolet coherence tomography}

\hypersetup{
pdftitle={Non-destructive depth reconstruction of Al-Al\textsubscript{2}Cu layer structure with nanometer resolution using extreme ultraviolet coherence tomography},
pdfsubject={physics.optics},
pdfauthor={Johann Jakob Abel, Jonathan Apell},
pdfkeywords={First keyword, Second keyword, More},
}

\begin{document}
\maketitle

\begin{abstract}
Non-destructive cross-sectional characterization of materials systems with a resolution in the nanometer range and the ability to allow for time-resolved in-situ studies is of great importance in material science. Here, we present such a  measurements method, extreme ultraviolet coherence tomography (XCT). The method is non-destructive during sample preparation as well as during the measurement, which is distinguished by a negligible thermal load as compared to electron microscopy methods. Laser-generated radiation in the extreme ultraviolet (XUV) and soft x-ray range is used for characterization. The measurement principle is interferometric and the signal evaluation is performed via an iterative Fourier analysis. The method is demonstrated on the metallic material system Al-Al\textsubscript{2}Cu and compared to electron and atomic force microscopy measurements. We also present advanced reconstruction methods for XCT which even allow for the determination of the roughness of outer and inner interfaces.
\end{abstract}


\section{Introduction}

Transmission electron microscopy is undeniably one of the most important tools for microstructural analysis in materials science. It provides information on key aspects such as local composition, crystal structure (including phase identification) and crystal orientation \cite{williams_transmission_2009}. A number of imaging and diffraction modes are available for this purpose, providing a large number of contrasts and structural information. In addition, spectroscopy methods such as energy dispersive X-ray spectroscopy or electron energy loss spectroscopy provide detailed information about the composition \cite{williams_transmission_2009}. These techniques are indispensable for pursuing scientific and technological questions in materials science. However, there are also limitations due to specimen damage during sample preparation and measurement, including heat deposition and amorphization  during ion milling \cite{mayer_tem_2007, viguier_heating_2001} as well as knock-on damage, radiolysis, and specimen heating during the measurement \cite{williams_transmission_2009, egerton_radiation_2019, neelisetty_electron_2019, jiang_electron_2015}.

For \textit{in-situ} measurements, high spatial and high temporal resolution are in conflict with each other. For example, it is not possible to measure the velocity of interfaces during rapid phase transformations and to conclude on the thermodynamic state of the interface, as relevant time and length scales are on the order of $1\,\mathrm{\upmu s}$ and $2-10\,\mathrm{nm}$, respectively \cite{gamsjager_kinetics_2007, rettenmayr_modeling_2008, mathiesen_situ_2012, thi_preliminary_2003}. The limitation concerns both solid/liquid phase transformations with pronounced jumps in the chemical potential, triggered by high temperature or concentration gradients, and solid state transformations with partial or complete interfacial control (martensitic and massive transformation).

Materials characterization with extreme ultraviolet (XUV) radiation is an excellent way to overcome some of these limitations. It provides information on local parameters, \textit{e.g.}, chemical and stoichiometric compositions \cite{yakunin2014combined}, depth structure \cite{le2010development}, surface and interface roughness \cite{sertsu2015analysis} and thus allows conclusions on phases present and their spatial extension. Usually, such measurements are performed using XUV radiation at synchrotron facilities. However, laser-based XUV sources increasingly enable characterization methods established at synchrotrons to be transferred to university-scale laboratories. The process of high harmonic generation (HHG) is a key technology in this respect. Our method of XUV coherence tomography (XCT) is based on white light interferometry and utilizes the broad bandwidth of HHG for non-destructive nanoscale subsurface imaging.

With the development of XUV microscopy in general and laser based XCT in particular, entirely new possibilities become available. The methodology combines a depth resolution in the few nanometer range \cite{Fuchs2017,wachulak2018optical,skruszewicz2021coherence} with the option of extremely high temporally resolved measurements in the pico- to femtosecond range. At the same time, heat insertion is minimal, as a power of only up to $50\,\mathrm{W/m^2}$ is used.

We describe and apply the method of XCT for non-destructive characterization of a metallic layer system composed of an intermetallic phase (Al\textsubscript{2}Cu) and a solid solution ($\alpha$-Al). In the current state of development, the focus is on the achievable material contrast and the quantification of sample depth profiles with nanometer resolution. We further report on the reconstruction of layer thickness and roughness parameters of present native surface and interface oxides, using a novel reconstruction algorithm based on a step wise optimization routine. Retrieved sample parameters from non-destructive XCT measurements are validated and compared to atomic force microscopy (AFM) as a surface sensitive method and transmission electron microscopy (TEM) as an invasive characterization method.

\section{Materials and Methods}

\subsection{Extreme ultraviolet coherence tomography (XCT)}

\subsubsection{Principle of XCT} 

\begin{figure}[ht]
\centering\includegraphics[width=0.45\linewidth]{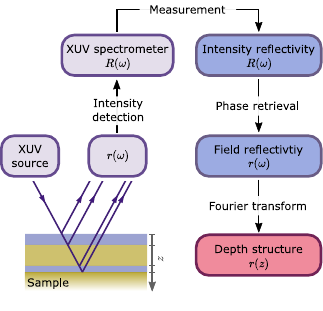}
\vspace{-0.3cm}
\caption{Principle of depth profile reconstruction with XCT based on common-path interferometry. Scattered radiation from different interfaces interferes with each other, which leads to modulations in the recorded reflectivity spectrum. The frequencies of spectral modulations carry the structural depth information of the sample, which can be extracted by Fourier-based algorithms.}
\label{fig:common-path}
\end{figure}

Extreme ultraviolet coherence tomography (XCT) is a non-destructive tomographic imaging technique, well suited for depth profile reconstruction of nanoscale layered samples with high resolution \cite{Fuchs2016,wiesner_oe_22}. It is based on the principles of white light interferometry and uses a Fourier-domain common path interferometric setup \cite{Fuchs2012}, which is depicted in Figure \ref{fig:common-path}. Scattered light from different interfaces interferes in the detected reflection beam and causes spectral modulations, which carry the structural depth information.

The complex field reflectivity response $r(\omega)$ of a layer structured sample can be expressed by the sum of $n$ independent interface reflectivities $r_j$ located at depths $z_j$ and reads \begin{equation}
   r(\omega)=\sum_{j=1}^n r_j(\omega)\cdot\exp\left[2ik_z(\omega)z_j\right],
\label{eq:r}
\end{equation} 
where the real part of the wave vector component $k_z(\omega)$ approximates the dispersion of the dominant component within the sample. The axial sample structure $r(z)$ can be reconstructed by the Fourier transformation of \eqref{eq:r}. When using a common-path interferometer, the phase information of the response is lost and thus only the intensity reflectivity $R(\omega)=|r(\omega)|^2$ is measured. However, reconstruction of the field response from $R(\omega)$ is possible using a three-step, one-dimensional, iterative phase retrieval algorithm to recover the lost phase information and subsequently determine the unique sample response $r(z)$ \cite{Fuchs2017}. 

The maximum achievable depth resolution $\Delta z$ depends mainly on the bandwidth of the radiation source used and the XUV transmission bandwidth of the sample's dominant material. It is given by
\begin{equation}
    \Delta z\,\sim\,\frac{\lambda_0^2}{\Delta \lambda}\,\sim\,\Delta E.
\end{equation}
Thus, broadband short wavelength radiation in the spectral region of XUV and soft x-rays enables imaging with nanometer resolution \cite{skruszewicz2021coherence}. Additionally, the refractive index in this spectral region is dominated by characteristic atomic levels of inner K-, L- and M-shells. Therefore, the individual reflectivities of different materials provide a unique elemental material contrast. However, since this spectral region is dominated by strong absorption of most elements, XCT is restricted to sample inspection under vacuum conditions, a limited penetration depth of a few micrometers, and spectral transmission windows depending on the absorption of the materials in the sample.

\subsubsection{Experimental XCT setup}

\begin{figure}[ht]
\centering\includegraphics[width=0.7\linewidth]{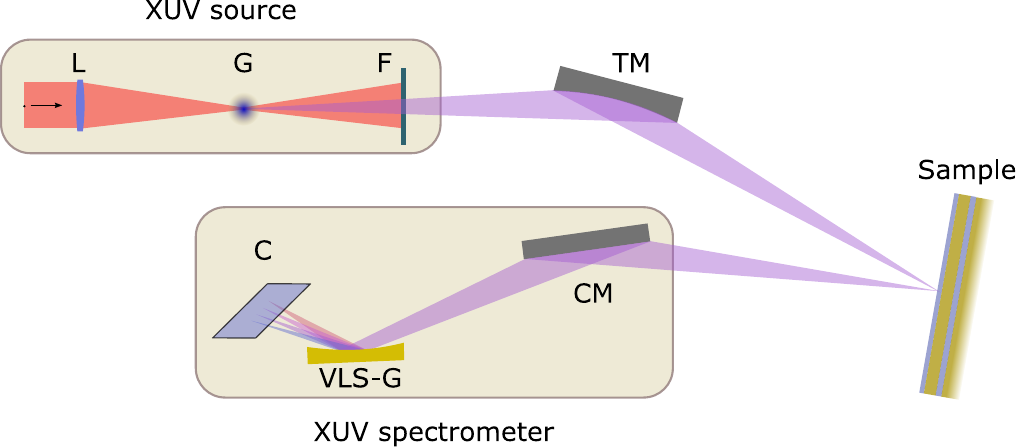}
\caption{Basic XCT setup consisting of a high harmonic XUV source, the toroidal focusing mirror TM, the sample and an imaging XUV spectrometer.}
\label{fig:setup}
\end{figure}

The experimental method rests on recording reflection spectra, which have been measured using a laser-based broadband XUV beamline \cite{Nathanael2019}. An overview of the used setup is shown in Figure \ref{fig:setup}. The high-harmonic radiation between 30 and 100\,eV is generated by focusing an intense ultrashort laser pulse ($E_P$\,=\,2\,mJ, $\tau$\,=\,35\,fs, $\lambda$\,=\,1300\,nm) into an argon gas jet (G) by means of a lens with $f$\,=\,30\,cm focal length (L). Residual light of the driving laser is removed by a 200\,nm thin Al filter foil (F), while the transmitted XUV radiation is focused by a 1-m focal length toroidal mirror (TM), thus illuminating a spot size of around 50\,µm on the sample under investigation. The use of Al for filtering limits the useful transmission window to $<$\,72.6\,eV, which is connected to the fundamental L-absorption. Consequently, an Al dominated sample itself is restricted to be measured within this window. The used bandwidth in the experimental data supports an axial (i.e. depth) resolution of 35\,nm.
Finally, the radiation is reflected at the sample at an angle of incidence of 15$^\circ$ and is subsequently detected using a high-resolution imaging XUV spectrometer \cite{Wunsche2019}. It includes a cylindrical focusing mirror (CM), a variable line space grating (VLS-G) and a CCD camera (C). In order to cover the full spectral window, quasi-continuous XUV radiation is generated by fast wavelength shifting of the driving laser field during the sample exposure \cite{Wunsche2017}. 

An additional 20\,nm thin Si$_3$N$_4$-membrane, which is placed directly in front of the sample is used to split the incoming beam in order to enable a simultaneous reference measurement of the source spectrum \cite{Abel2022}. This procedure provides precise data on absolute broadband reflectivity values, which is essential for the subsequent phase retrieval and sample reconstruction.

\subsubsection{Signal analysis and sample reconstruction}

\begin{figure}[ht]
\centering\includegraphics[width=1\linewidth]{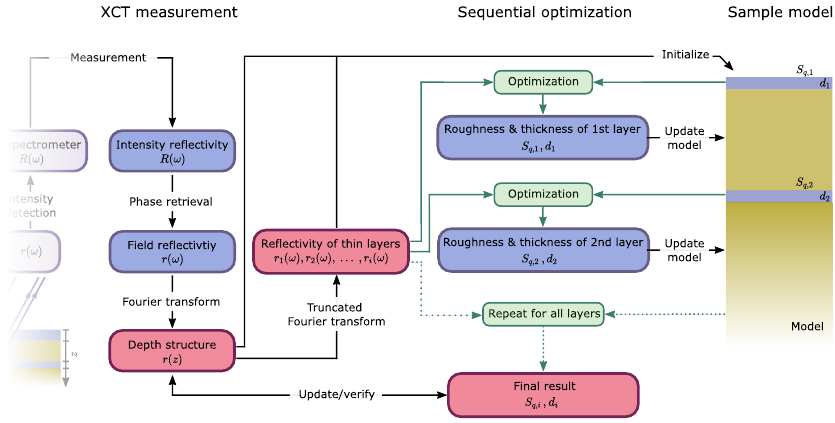}
\caption{Flowchart of structural sequential sample reconstruction based on XCT measurements and simulations from a sample model. The left edge of the figure merges into Figure \ref{fig:common-path}.}
\label{fig:algorithm}
\end{figure}

An overview of the further data processing after the measurement is shown in Figure \ref{fig:algorithm}. The absolute reflectance of the sample $R(\omega)$ is determined from the simultaneously recorded and calibrated spectra of the reference membrane and the sample and serves as a starting point for further sample reconstruction.

First, the spectral field response $r(\omega)$ of the sample is recovered using one-dimensional phase retrieval algorithms \cite{Fuchs2017}. The iteratively determined solution with a phase that minimizes the error to the measured spectrum allows the unambiguous assignment of the sample's depth structure using a Fourier transform. At this point, the depth profile of the sample under investigation is unambiguously revealed for the first time.

Additionally, one can extract material specific spectral information of the individual interfaces. This is realized by truncated Fourier transform of $r(z)$. The according interfaces are selected by sequential spatial filtering of $r(z)$ before the Fourier transform, which reveals $r_j(\omega)$ of the addressed interface \cite{Wiesner2021}.

The depth structure together with the additional spectral information for each interface can further be used for model-based fitting of structural parameters such as thickness and interface roughness for each layer. The respective procedure starts with the measured depth profile. Information on material composition of the sample  is included based on prior knowledge or from the just determined characteristic spectral reflectivities $r_j(\omega)$. From the sample model, which includes the materials, thicknesses $d_i$ as well as the surface and the interface roughnesses $S_{q,i}$, the field reflectivity is calculated using the transfer matrix formalism \cite{saleh2019fundamentals} together with  Nevot-Croce factors \cite{nevot1980caracterisation}  and tabulated XUV dispersion data from \cite{henke1993x}. This model-based data is then processed in exactly the same way as the measurement data and allows an optimization based on minimizing an error function between model-based data and the measurement.

Structural properties such as layer thicknesses $d_i$ and roughnesses $S_{q,i}$ of the sample are determined sequentially from top to bottom in the optimization presented here. This means that initially only the properties of the first layer are determined. Subsequently, the reconstructed parameters ($d_1, S_{q,1}$) of the first layer are used to improve the sample model. On this basis, the next layer beneath is optimized. Since dispersion, absorption, and scattering losses due to the thickness and roughness of the layer above are already known from the first optimization step, only the parameters of the lower layer need to be further optimized. The procedure is repeated until the lowest layer has been addressed. In this way the sample is reconstructed step by step from top to bottom.

\begin{figure}[ht]
\centering\includegraphics[width=0.43\linewidth]{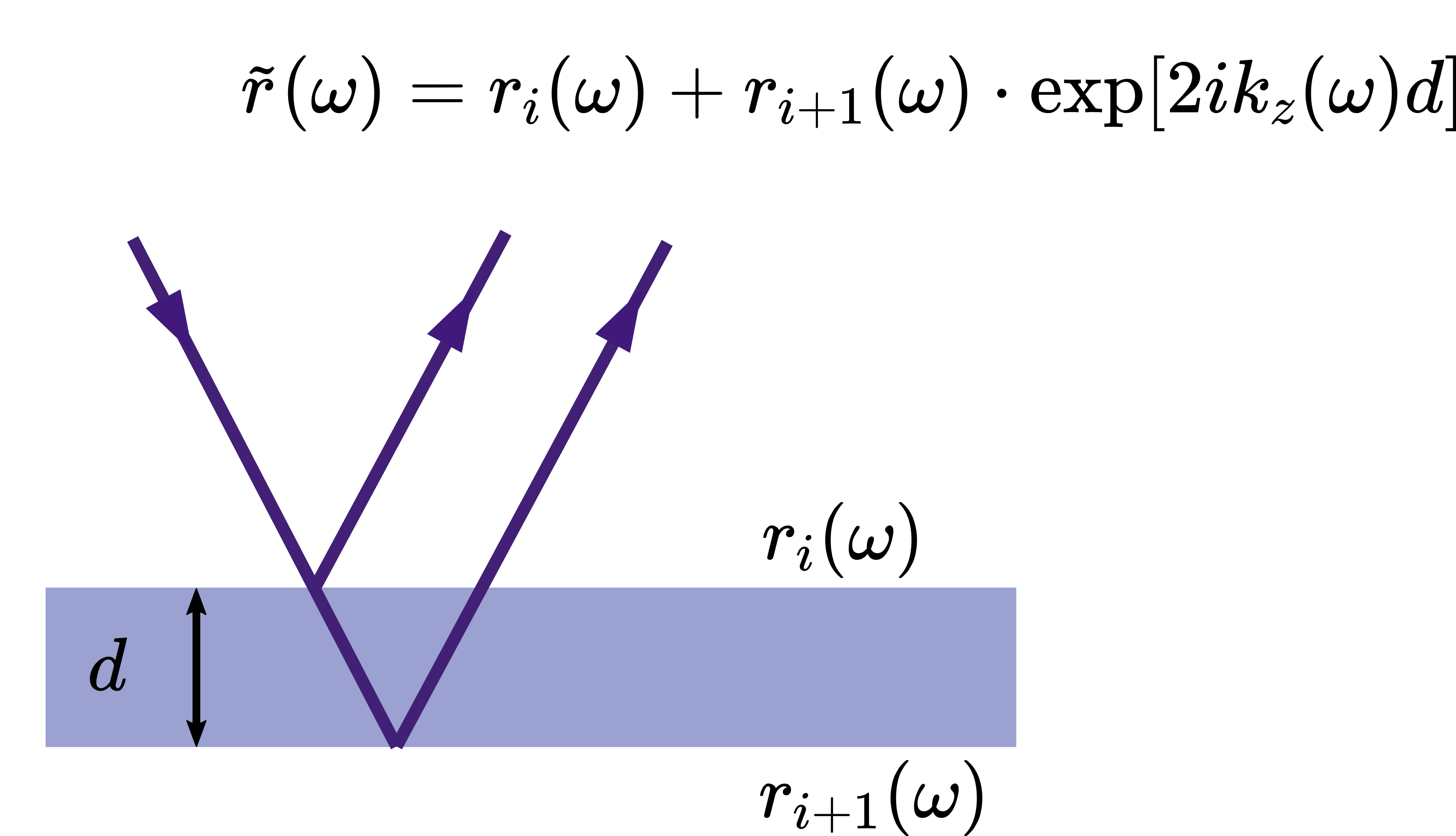}
\caption{The complex-valued field reflectivity $\tilde{r}_{\text{j}}(\omega)$ of a thin layer. The phase strongly depends on the layer thickness $d$, which allows for the characterization of ultra thin layers even below the limit of depth resolution.}
\label{fig:r}
\end{figure}

Employing an optimization of the sample model to fit our measurements additionally allows for the detection of very subtle changes in layer thickness far below the nominal (i.e. model-free) depth resolution of XCT. The two interfaces of a layer with a thickness below the depth resolution of XCT can not be imaged, only a superposition $\tilde{r}_{\text{j}}(\omega)$ of both interface reflectivities is reconstructed by the phase-retrieval algorithm.  However, using the knowledge about the existence of a thin layer, it is possible to also reconstruct its thickness. A variation in the range of single nanometers has a significant effect on $\tilde{r}_{\text{j}}(\omega)$, since the thickness has a fairly strong influence on the phase difference as shown in Figure \ref{fig:r}. Ultimately, the algorithm achieves sub-nanometer accuracy for the layer thickness.

The sequential optimization routine presented here is fundamentally different from a direct multidimensional fit of the reflected spectrum. The fitting procedure based on distinct spectral functions $r_j(\omega)$ is very advantageous since the spatial restrictions during the filtering process strongly reduce numerical instability. In the picture of frequency space, the truncated Fourier-transform acts as a sharp frequency filter and only the modulation that corresponds to the related interface under investigation for the single optimization step is further evaluated. Thus, the presented reconstruction offers the advantage to optimize physically measured quantities by single sequential optimization steps.

\subsection{Al-Al\textsubscript{2}Cu sample preparation}

An Al-Al\textsubscript{2}Cu layer system was prepared for investigation with XCT. Al$_2$Cu is an incongruently melting phase whose preparation as a pure bulk material is only possible by directional solidification in a temperature gradient \cite{Lffler2014} or by solidification in the metastable \cite{Prince2020, Fang2022}. In this work, an induction levitation furnace with a cold-wall crucible that allows the suppression of the primary phase was used. The Al$_2$Cu substrate was prepared from raw materials of $\geq$99.99\% purity. A 20\,g ingot with the composition Al-33.3Cu (in at.\%) was inductively melted in a cold-wall crucible in Ar atmosphere. The ingot was cut into samples with dimensions of $10\,\mathrm{mm} \times 20\,\mathrm{mm} \times 2\,\mathrm{mm}$ with a high-speed saw. Sample surfaces were ground with SiC abrasive paper successively up to 1200 grit and polished to a mirror finish with $9\,\mathrm{\upmu m}$, $3\,\mathrm{\upmu m}$, $1\,\mathrm{\upmu m}$, and $0.25\,\mathrm{\upmu m}$ polycrystalline diamond suspensions.
The polished Al$_2$Cu sample surfaces were coated with a $\sim290\,\mathrm{nm}$ thick Al layer using DC magnetron sputtering in a redesigned Perkin-Elmer 4480 Sputter Deposition Systems equipped with three Delta\textsuperscript{TM} cathode positions. The argon gas pressure in the deposition chamber was $2.5 \cdot 10^{-2}$\,mbar and a background pressure of $2 \cdot 10^{-7}$\,mbar was used. The sputtering power was kept at $1.5$\,kW, resulting in a deposition rate of $0.34$\,nm/s. After sputter coating, a final polishing step was carried out with $50\,\mathrm{nm}$ alkaline colloidal silica suspension (Cloeren Technology, Germany) in order to achieve a surface root mean square roughness below $5\,\mathrm{nm}$ while preserving most of the Al layer.

\subsection{Sample characterization with atomic force microscopy and transmission electron microscopy}

In order to compare the results obtained from the XCT reconstruction with results obtained from established characterization methods, additional characterization of layer thickness and roughness was carried out. The root mean square roughnesses $S_q$ of the substrate before and after coating and polishing were determined in an area of $1\,\mathrm{\upmu m} \times 1\,\mathrm{\upmu m}$ using atomic force microscopy (AFM, Bruker Dimension Icon). The AFM was operated in tapping mode utilizing single crystalline silicon probes with a nominal tip radius of $7\,\mathrm{nm}$. \\
Cross sections of the Al-Al\textsubscript{2}Cu samples were prepared for transmission electron microscopy (TEM) using a dual beam focused ion beam system (FEI Helios NanoLab 600i). The sample surface was covered with protective Pt before the cross section preparation to minimize damage during sample preparation. The average thicknesses of the Al layer and the observed oxide layers were determined from 10 measurements along the TEM lamella. High resolution transmission electron microscopy (HRTEM) and scanning transmission electron microscopy (STEM) imaging was carried out on the Al-Al\textsubscript{2}Cu sample (JEOL NEOARM 200 F). Nano beam electron diffraction (NBED) was used to record diffraction patterns of the substrate and the sputter-coated Al layer. Crystallographic phases were identified from NBED patterns and overlaid with corresponding calculated diffraction patterns using SingleCrystal 4.1.2 (CrystalMaker Software Ltd., Begbroke, UK) to verify the identified phases. STEM equipped with energy-dispersive X-ray spectroscopy (EDXS) was used to map the concentration distributions in the sample. EDXS concentration profiles were extracted from the STEM EDXS maps.

\section{Results}

\subsection{Structural sample reconstruction with XCT}

\begin{figure}[ht]
\centering\includegraphics[width=0.7\linewidth]{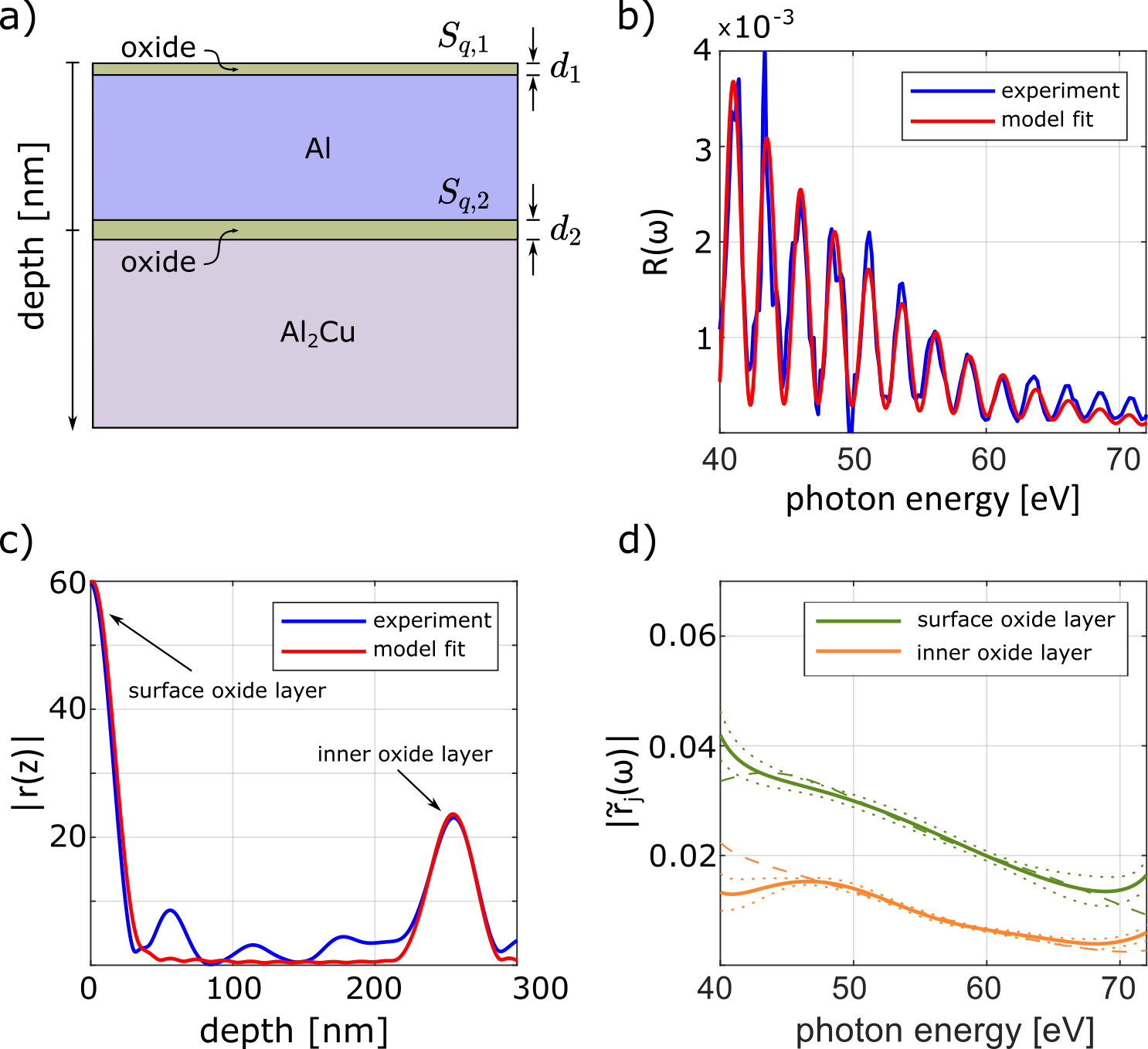}
\caption{(a) Structural model of the sample's depth profile. (b) Measured (blue) and calculated reflectivity curve from the reconstructed sample model (red). The modulations encode the buried interfaces. (c) The sample's depth profile $r(z)$ is reconstructed by a phase retrieval algorithm applied to measured reflectivity data (blue). The calculated response (red) is obtained from reflection data using the final reconstructed sample model. (d) Truncated Fourier transform retrieves the effective spectral reflectivity of the surface (green) and the buried oxide layers (orange). Dashed lines are measured data. The solid lines represent calculated data from the sample model and dotted lines display their corresponding error intervals.}
\label{fig:result}
\end{figure}

XCT measurement, data processing, and final structural sample reconstruction were performed as described in section 2.3. The structural model of the sample is shown in Figure \ref{fig:result}a. In addition to the already specified Al$_2$Cu and Al layer, the existence of a surface oxide layer and a buried oxide layer between Al$_2$Cu and Al was considered. Al$_2$O$_3$ is the equilibrium oxide for both phases, i.e. native Al$_2$O$_3$ layers (typically less than 10\,nm thick) are expected to form almost instantaneously when in contact with air \cite{Seyring2021}. Prior to the measurement and sample reconstruction, the thicknesses of the oxide layers, just like that of the metallic layers, is unknown, only their possible existence was postulated.

The measured reflection spectrum between 40 and 72 eV is shown in blue in Figure \ref{fig:result}b. The spectral width of 32 eV provides an axial resolution of 35 nm. Thus, the depth of the interface (including the possible oxide layer) can be reconstructed by a Fourier transform. For reconstruction of the thicknesses of the oxide layers, we used the optimization procedure described above based on the sample model, in which the thicknesses as well as layer roughnesses are included as free parameters. 

For the present sample, this works as follows: A phase retrieval is used to determine the absolute values of the samples field response $|r(z)|$, which is shown in blue in Figure \ref{fig:result}c. This forms the starting point for the determination of material-specific spectral reflectivity curves of the individual interfaces within the sample. For this purpose, the respective peak is first selected and filtered between the first minima on the left and on the right in spatial domain. While the surface peak can be Fourier transformed directly and results in the effective surface reflectivity $\tilde{r}_{\text{1}}(\omega)$ (dashed green curve in Figure \ref{fig:result}d), the depth signal of the buried oxide needs to be shifted to zero first. As a result, the solid orange curve for the effective interface reflectivity $\tilde{r}_{\text{2}}(\omega)$ is obtained.

Theoretical intensity and field reflectivities from a specific sample model structure are calculated using the transfer matrix formalism \cite{saleh2019fundamentals} and XUV dispersion data from \cite{henke1993x}. Subsequently, the calculated field reflectivity obtained from the model is processed in the same way as the measured data including spatial filtering and Fourier transform. This allows a direct comparison between the XCT measurement and calculations from the sample model.

Starting from the surface, the sample structure is reconstructed and the sample model is updated step by step. In the first step, only the surface roughness $S_{q,1}$ and surface oxide thickness $d_1$ are optimized, by minimizing the deviation between measured and simulated $\tilde{r}_{\text{1}}(\omega)$ of the surface. The optimization was performed in a spectral range from 47 to 65\,eV, since the edges of the spectral signal feature larger deviations due to strong spatial filtering. The optimization yields a surface roughness of $S_{q,1}=2.4\pm 0.2$\,nm and an oxide thickness of $d_1=2.1\pm 0.3$\,nm. These values are now fixed in the model during further optimization. The resulting calculated effective surface reflectivity $\tilde{r}_{\text{1}}(\omega)$ of the optimized surface oxide in the sample model is shown as a solid green line in Figure \ref{fig:result}d.

Subsequently, the thickness $d_2$ and roughness $S_{q,2}$ of the buried oxide layer is optimized. Therefore, the structural sample model is refined by including the previously reconstructed values for the surface oxide thickness and roughness. Again the deviation between model based calculations and the measurement is minimized to recover $d_2$ and $S_{q,2}$. The sum of oxide thickness and Al layer thickness is kept constant, since the absolute depth is already known from $|r(z)|$ (see Figure \ref{fig:result}c). A thickness of $d_2=3.7\begin{smallmatrix}+1.4\\-1.0\end{smallmatrix}$\,nm and roughness of $S_{q,2}=3.5\begin{smallmatrix}+0.1\\-0.2\end{smallmatrix}$\,nm is obtained for the buried interface oxide and results in the solid orange simulated curve in Figure \ref{fig:result}d. The thickness of the Al layer is determined to be $230.5\begin{smallmatrix}+0.7\\-0.9\end{smallmatrix}$\,nm.

\subsection{Cross section analysis with TEM}

\begin{figure}[!htbp]
    \centering
    \includegraphics[width=140mm]{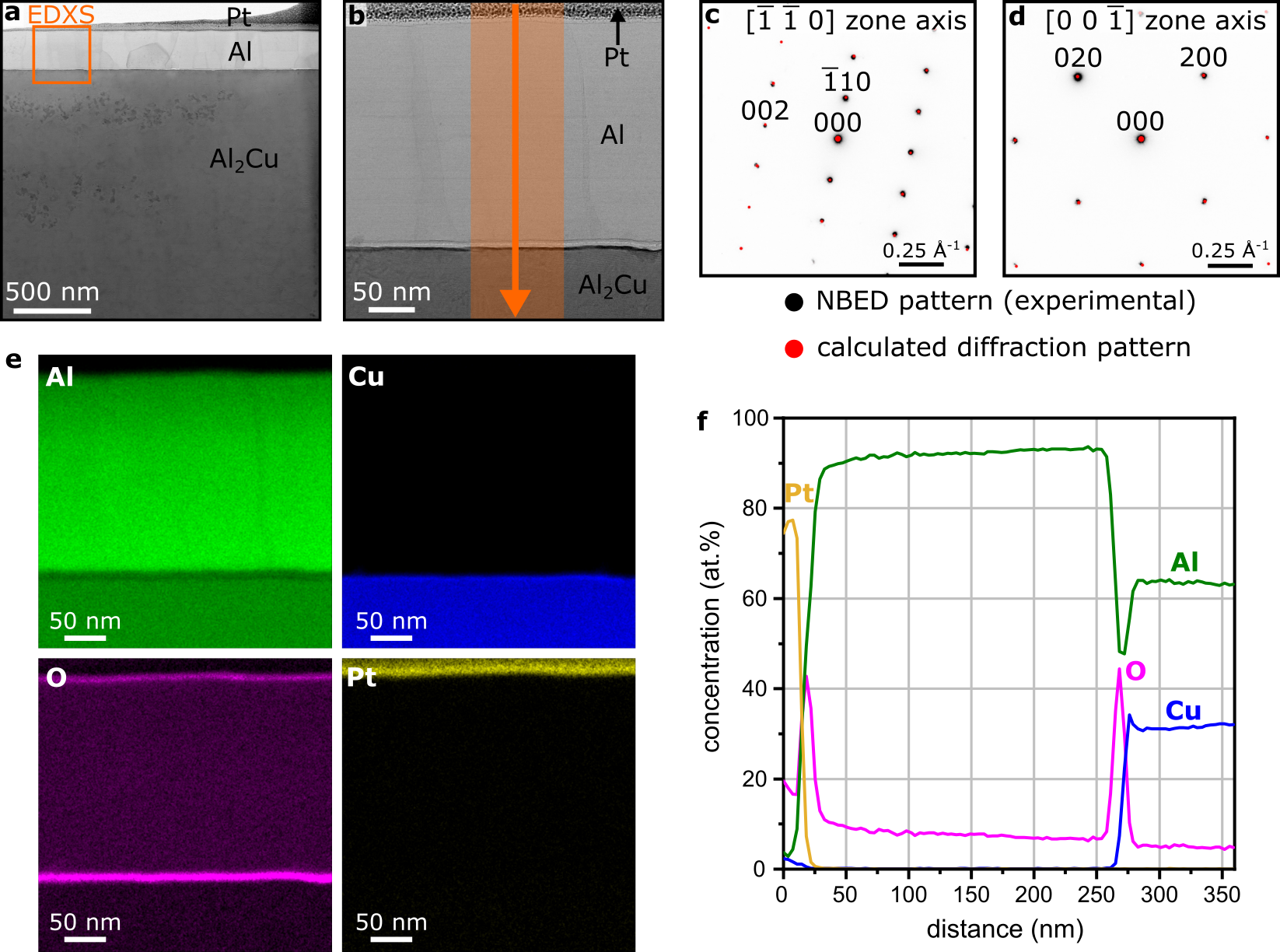}
    \caption{TEM results of the Al$_2$Cu-Al sample. (a,b) STEM bright field images showing the Al-Al$_2$Cu-Al layer system. NBED patterns recorded from the Al$_2$Cu substrate (c) and the Al layer (d). The experimental and calculated diffraction patterns according to the phase identification are overlaid. (e) STEM EDXS maps of the region marked in (a) and shown in (b). (f) Extracted EDXS profile along the orange shaded region in (b).}
    \label{fig:results_STEM_EDX}
\end{figure}

Characterization using TEM allows for independently determining layer thicknesses and compositions with high resolution to verify the XCT reconstruction results. STEM bright field images show the Al$_2$Cu substrate below the sputter coated Al layer (Figure \ref{fig:results_STEM_EDX}a and b). Dark regions in the Al$_2$Cu substrate are a Cu-rich phase, $\eta$-AlCu, that formed in the last steps of the solidification process due to a local solute enrichment. The Al layer has a homogeneous thickness across the width of the TEM lamella. The Al layer thickness excluding oxide layers is measured to be $233 \pm 2\, \mathrm{nm}$. Phase identification using NBED of the Al$_2$Cu substrate confirms the tetragonal Al$_2$Cu phase with space group \textit{I4/mcm} (Figure \ref{fig:results_STEM_EDX}c). The Al layer is confirmed to be in face centered cubic structure in space group \textit{Fm-3m} (Figure \ref{fig:results_STEM_EDX}d). Calculated diffraction patterns fit well with the experimentally recorded NBED patterns. STEM EDXS analysis of the region marked in Figure \ref{fig:results_STEM_EDX}a and given in Figure \ref{fig:results_STEM_EDX}b reveals oxide layers at the Al$_2$Cu/Al interface and on the Al surface (Figure \ref{fig:results_STEM_EDX}e). The extracted EDXS profile along the orange shaded area in Figure \ref{fig:results_STEM_EDX}b confirms a layer of pure Al on an Al$_2$Cu substrate. The oxide layers at the surface and the Al$_2$Cu/Al interface are Al-rich and contain no Cu. The low O content observed in the Al layer and Al$_2$Cu substrate likely stems from native oxide layers forming on the top and bottom of the TEM lamella during thepreparation process.

\begin{figure}[!htbp]
    \centering
    \includegraphics[width=140mm]{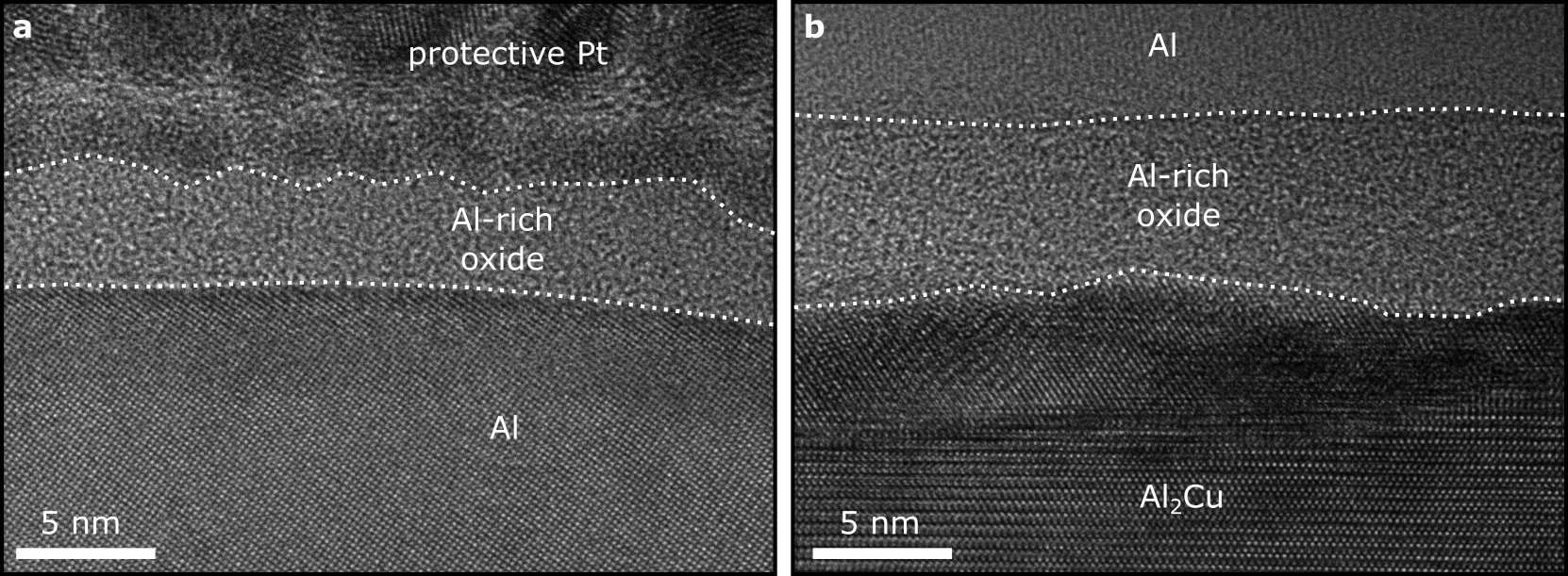}
    \caption{HRTEM images of the oxide layer on the Al surface (a) and at the Al$_2$Cu/Al interface (b).}
    \label{fig:results_HRTEM}
\end{figure}

In order to characterize the thin oxide layers at the surface and Al$_2$Cu/Al interface in detail, HRTEM was carried out. Both Al-rich oxide layers are found to be amorphous (Figure \ref{fig:results_HRTEM}). The thickness of the oxide that formed on the sample surface (Figure \ref{fig:results_HRTEM}a) is measured to be $d_1^{\text{TEM}} = 4.1 \pm 0.3\, \mathrm{nm}$. The oxide at the Al$_2$Cu/Al interface (Fig. \ref{fig:results_HRTEM}b) is slightly thicker at $d_2^{\text{TEM}} = 6.0 \pm 0.7\, \mathrm{nm}$. 

\subsection{Roughness analysis of substrate, coated and polished samples with AFM}

\begin{figure}[!htbp]
    \centering
    \includegraphics[width=140mm]{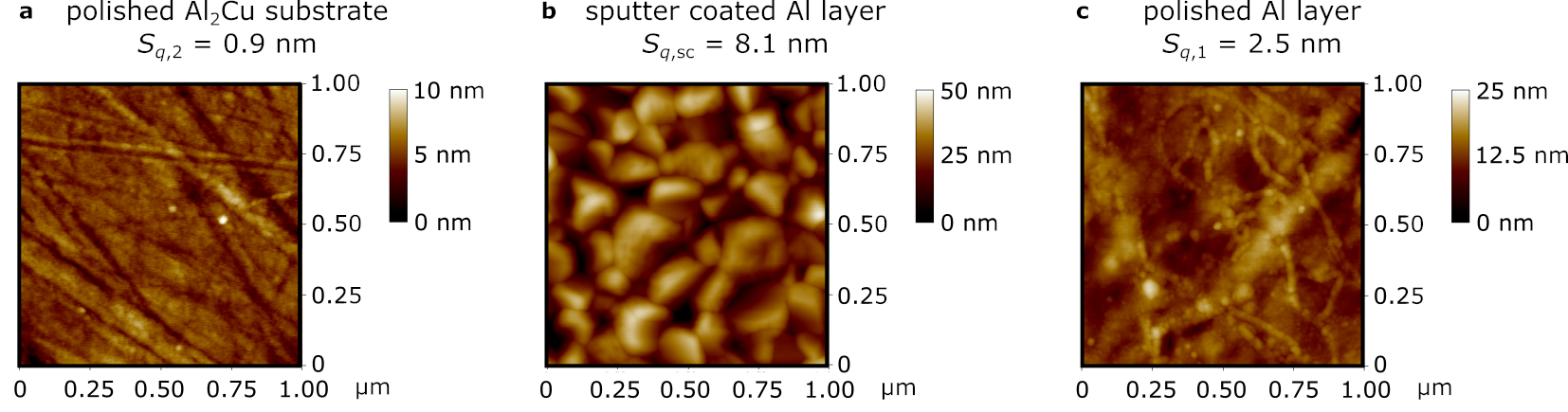}
    \caption{AFM scans and measured RMS roughnesses $S_q$ for the Al$_2$Cu substrate (a), the sputter-coated Al surface (b), and the polished Al surface (c).}
    \label{fig:results_AFM}
\end{figure}

Characterization using AFM allows for independent determination of surface roughness of the sample before and after sputter coating. The AFM scan of the Al$_2$Cu substrate before coating shows a surface with visible polishing grooves and a low roughness of $0.9\,\mathrm{nm}$, corresponding to the roughness of the inner oxide layer $S_{q,2}$ (Figure \ref{fig:results_AFM}a). After sputter coating of the 290-nm Al layer, the Al surface is made up of individual grains with a significantly larger surface roughness of $S_{q,\mathrm{sc}} = 8.1\,\mathrm{nm}$ (Figure \ref{fig:results_AFM}b). In order to enable an XCT measurement on the Al$_2$Cu-Al sample, a further polishing step was carried out as described above. After polishing, the sample surface is much smoother with an achieved final roughness of $2.5\,\mathrm{nm}$, corresponding to the surface roughness $S_{q,1}$ (Figure \ref{fig:results_AFM}c). 
\vspace{0.5cm}

\section{Discussion}

The XCT method presented in this paper enables precise, non-destructive characterization of nanometer-scale layered systems. A complete reconstruction of the Al-Al\textsubscript{2}Cu layer system including thickness and roughness values has been demonstrated. First, the layer thicknesses and interfacial roughnesses determined by XCT are compared with results obtained by TEM and AFM measurements (section 4.1). Additionally, the error propagation and the confidence intervals of thickness and roughness values are discussed (section 4.2). 
Specific requirements for the surface quality of the sample under investigation and their influence on the sample reconstruction are discussed in section 4.3. Finally, the presented XCT method is compared to established characterization methods and possible applications are highlighted (section 4.4).

\subsection{Comparison of layer thickness and roughness values obtained by XCT reconstruction, TEM, and AFM}

A comparison of the layer thickness and roughness values attained by the XCT measurement and subsequent reconstruction with the values measured using TEM and AFM is shown in Table \ref{tab:disc_comparison}. The Al layer thicknesses as determined separately by XCT and TEM show very good agreement. Slight deviations in the measured thickness may result from slight lateral variations in Al layer thickness on the sample as both XCT ($\sim50\,\mathrm{\upmu m}$ spot size) and TEM ($\sim10\,\mathrm{\upmu m}$ lamella width) measure locally. However, lateral scanning with XCT to investigate variations in thickness and roughness over the sample surface is straight forward. 

Both XCT and TEM confirm the buried oxide thickness $d_2$ to be larger than the surface oxide thickness $d_1$. The buried oxide thickness $d_2$  measured by TEM is close to the upper error limits of $d_2$ obtained from the XCT reconstruction. Differences in the measured surface oxide thickness $d_1$ can arise from effects during preparation of the TEM lamella, \textit{e.g.}, due to the Pt layer deposition on the sample surface during FIB preparation. Similar to the Al layer thickness, slight lateral variations in oxide layer thicknesses may also contribute to differences in the values measured with different techniques.

Surface roughness values $S_{q,1}$ determined by XCT and AFM are in excellent agreement. The substrate roughness $S_{q,2}$ as measured with AFM is significantly lower than the roughness of the buried oxide layer $S_{q,2}$ as reconstructed with XCT.
This difference can result from the formation of a transition layer between the substrate (Al\textsubscript{2}Cu) and the thin film material (Al) during the sputter coating. This is supported by the TEM results, as a thin transition layer is visible as a dark contrast in Figure \ref{fig:results_STEM_EDX}b and below the Al-rich oxide in Figure \ref{fig:results_HRTEM}b. This may lead to an increased buried interface roughness in XCT measurements when compared to the substrate roughness before sputter coating. In addition, lateral variations in the substrate roughness due to polishing grooves may influence the roughness results obtained by AFM, as the $1\,\mathrm{\upmu m} \times 1\,\mathrm{\upmu m}$ measurement area is small compared to the XCT spot size.

\begin{table}[!htbp]
    \centering
    \begin{tabular}{l l c c}
    measured parameter (nm) & XCT & TEM & AFM\\ 
    \hline
    \\
    Al layer thickness & $230.5\begin{smallmatrix}+0.7\\-0.9\end{smallmatrix} $ & $233 \pm 2.0$ & - \\
    surface oxide thickness $d_1$ & $2.1\pm 0.3$ & $\,4.1 \pm 0.3$ & - \\
    buried oxide thickness $d_2$ & $3.7\begin{smallmatrix}+1.4\\-1.0\end{smallmatrix} $ & $\,6.0 \pm 0.7$ & - \\
    surface roughness $S_{q,1}$ & $2.4\pm 0.2$ & - & $2.5$ \\
    substrate/buried oxide roughness $S_{q,2}$ & $3.5\begin{smallmatrix}+0.1\\-0.2\end{smallmatrix} $ & - & $0.9$ \\
    \\
    \end{tabular}
     \caption{Layer thickness and roughness values attained by XCT, TEM, and AFM.}
    \label{tab:disc_comparison}
\end{table}

\subsection{Error estimation}

To estimate the error of the measured effective layer reflectivities $\tilde{r}_{\text{1}}$ and $\tilde{r}_{\text{2}}$, the measured reflection spectrum $R(\omega)$ is modified using different randomized spectral noise terms that have been characterized for the used setup in \cite{Abel2022}. For each spectrum from the obtained ensemble of spectra, phase reconstruction and truncated Fourier transform are performed and thus mean and standard deviation for $\tilde{r}_{\text{1}}$ and $\tilde{r}_{\text{2}}$ are determined (Fig. \ref{fig:result}d). Confidence intervals for the reconstructed layer parameters $d_j$ and $S_{q,j}$ are then obtained by the model-based optimization. However, since the optimization is done sequentially, the error propagation for each subsequent step needs to be included as well. It is mainly determined by the transmission of the already determined sample structure above the respective interface. In each optimization step $j$ the sets of parameters $d_{k}$ and $S_{q,k}$ with $k<j$ that cause maximal and minimal transmission are considered for further optimization. In this way the confidence interval for the optimization step $j$ also includes the error propagation from earlier steps. Larger confidence intervals are typically caused by a low depth signal strength of the respective buried layer. Therefore, besides the roughness of the interface itself, the detection limit of an extremely thin layer is limited by the transmission of the overlying layer system. This also implies a strong dependence on roughnesses of the overlying interfaces, as illustrated for example in section 4.3 using the surface roughness of the Al-Al\textsubscript{2}Cu layer system.

Whereas the 35\,nm depth resolution in the XCT measurement is limited by the spectral width of the measured reflection spectrum and describes the minimum detectable distance between two interfaces, the lower limit for a reliable layer thickness determination using the model-based reconstruction algorithm is basically given by the confidence interval of retrieved sample parameters. In the case of the surface oxide, the limit of a reliable thickness determination is roughly 0.5\,nm, wheras the limit for the interface oxide is already 1.0\,nm.

\subsection{Sample requirements}

In order to achieve successful XCT measurements and a valid sample reconstruction, a certain surface quality is an important prerequisite, since broadband XUV reflectivity measurements were performed in near-normal incidence geometry. Since the XUV illumination is focused, inhomogeneities on larger scales can be detected in advance and excluded. Homogeneous surface conditions need to be fulfilled over the illumination spot size of around 50\,µm. Smaller variations such as pores will influence the measured reflectivity and could possibly be detected. However, they will not be spatially resolved.

\begin{figure}[h!]
\centering\includegraphics[width=1\linewidth]{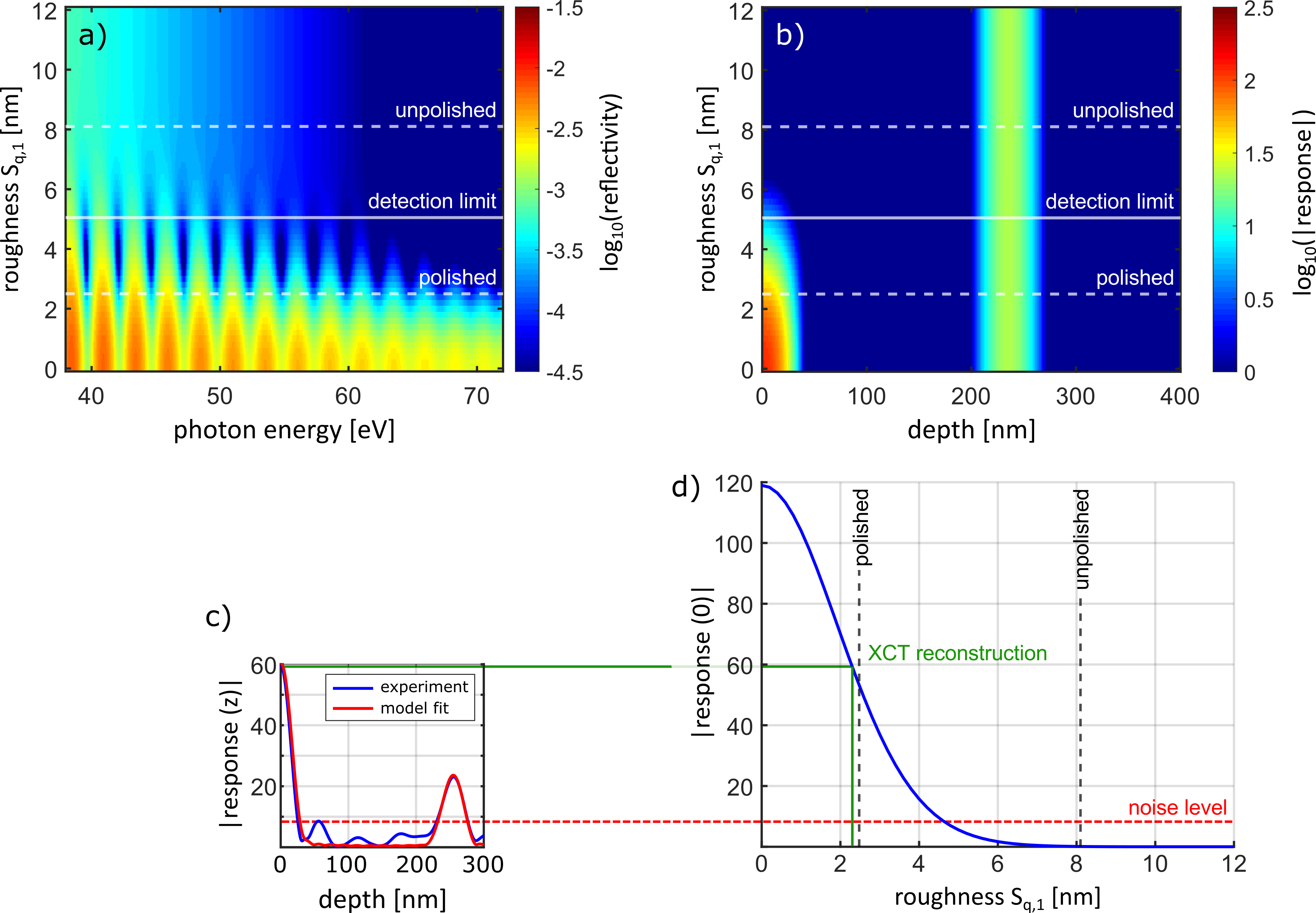}
\caption{Influence of the surface roughness $S_{q,1}$ on the XCT reconstruction based on simulations of the investigated sample structure. (a) An increasing surface roughness leads to a reduced modulation depth in the spectral reflectivity. (b) This is transferred to the sample response. (c) The detection limit due to observed noise level in the reconstructed response of experimental data specifies the detection limit due to the surface roughness, which needs to be smaller than $\sim$\,5\,nm RMS. (d) The amplitude of the surface response $r(0)$ as a function of the surface roughness is shown. The XCT reconstruction fits very well to simulated data including the measured surface roughness with AFM. Measurements with the unpolished substrate are not possible, since detected signal is below the noise level.}
\label{fig:roughness}
\end{figure}

The surface roughness of a layer-structured sample strongly influences its XUV reflectivity. Figure \ref{fig:roughness} illustrates the influence of the surface roughness on the course of the XUV field reflectivity (a) as well as the reconstructed depth structure (b). The reflectivity data in Figure \ref{fig:roughness}a are obtained from calculations based on the structure of the Al$_2$Cu-Al system investigated in the present work. The surface reflectivity was weighted by a Nevot-Croce factor to include roughness into the model \cite{nevot1980caracterisation}.

The Al$_2$Cu substrate had a low roughness of $S_{q,2}=0.9\,\mathrm{nm}$, enabling an XCT measurement with sufficient reflectivity. However, after sputter coating of the Al layer, the increased surface roughness of $S_{q,\mathrm{sc}} = 8.1\,\mathrm{nm}$ leads to a strongly decreased reflectivity, preventing a successful XCT measurement. This is consistent with the extracted detection limit due to the surface roughness of approximately $S_q = 5\,\mathrm{nm}$, which fits the $\lambda/4$ value of 20\,nm radiation (62\,eV). It is determined by the amplitude of residual noise in the depth response $|r(z)|$ obtained from the measurement as depicted in Figure \ref{fig:roughness}c and d. Only after the subsequent polishing step, which reduced the roughness to $S_{q,1} = 2.5\,\mathrm{nm}$, a strong reflectivity was observed and an XCT measurement was possible.

This illustrates a central point regarding sample design and preparation for XCT measurements for non-destructive depth reconstruction. Thicker sputtered layers are generally associated with a higher surface roughness \cite{lita_characterization_1999, elsholz_roughness_2005}, which has to be taken into account when preparing a layer system. When the initial surface roughness is too high, the sample can still be made suitable for XCT measurements with a subsequent polishing procedure. However, care has to be taken to choose a polishing procedure that decreases the surface roughness while preserving most of the sputter coated layer.

\subsection{Advantages of layer system characterization using XCT and comparison with established methods} 

In this study, we successfully applied the XCT method to characterize a layer system, evaluating both layer thicknesses and roughness down to a nanometer scale. In order to contextualize the novel method, a comparison with established characterization methods used for thin layers is presented, focusing on TEM, atom probe tomography (APT), and AFM.

The XCT method is non-destructive, allowing for further analysis of the same sample location using different techniques after an XCT measurement, thus facilitating a variety of analytical approaches. This non-invasive technique differs from traditional methods that often require destructive sample preparation. Whereas methods such as TEM or APT require ion milling with a focused ion beam \cite{mayer_tem_2007, gault_atom_2021}, a process that can potentially inflict significant damage to the specimen — including ion implantation or amorphization near the surface \cite{mayer_tem_2007} — XCT avoids these issues entirely. It maintains the surface of the sample without any artifacts that are generally associated with ion milling. 

In addition to artifacts from the sample preparation, beam damage during TEM analysis can result in knock-on damage, radiolysis, and localized specimen heating \cite{williams_transmission_2009, egerton_radiation_2019, neelisetty_electron_2019,jiang_electron_2015}. The extent of specimen heating during TEM analysis depends on numerous variables, particularly the thermal conductivity of the sample under investigation and the beam conditions utilized \cite{williams_transmission_2009}. Specimen heating is typically negligible for metallic materials with high thermal conductivity. However, it can be significant for materials with lower thermal conductivity like ceramics or polymers \cite{williams_transmission_2009}. Using XCT eliminates the risk of unwanted local specimen heating.

TEM and APT are known for their high spatial resolution, which facilitates the characterization of thin layers at the nanoscale. Although the depth resolution of XCT used in the present work is $35\,\mathrm{nm}$, the subsequent reconstruction algorithm allows for very accurate depth profile reconstruction comparable with TEM,
as discussed in section 4.1. The small illumination spot of XCT ($\sim50\,\mathrm{\upmu m}$) also enables the acquisition of localized data from small sample volumes, notably making it easier to laterally scan the sample surface. This approach simplifies the comprehensive 3D analysis of layered systems by scanning the sample over a 2-dimensional lateral grid \cite{Fuchs2017}.

Moreover, the XCT technique has several distinct advantages. It can evaluate layer thicknesses and discern roughness values, while also probing buried interface roughness, as highlighted in this study. Unlike AFM characterization, which is limited to the sample surface, XCT explores below, revealing unprecedented details.

The distinct material contrast provided by the XUV spectrum allows for precise material identification despite variations in composition. This enables the method to be applied to other material systems, including semiconductors and technologically relevant alloy systems such as Fe-, Cu-, Ni-, Mg- and Ti-base alloys. In particular, XCT is highly effective in characterizing thin oxide layers on metal or alloy substrates, indicating its potential as a key tool in the study of oxidation-resistant alloys. Such studies are increasingly significant, particularly in the investigation of the early stages of oxidation and thus thin oxide layers \cite{apell_microstructure_2023, yu_competition_2020, apell_early_2021}. Furthermore, a far-reaching characterization of different coating processes of thin layers becomes possible. It includes the simultaneous non-destructive determination of layer parameters such as thickness and roughness, which also allows conclusions to be drawn on the buried interface quality of processed
coated sample substrates.

Finally, the use of an HHG source provides the additional potential for ultrafast time-resolved XCT measurements, since the pulse duration of the produced XUV radiation can be on the femto- or even attosecond time scale.

\section{Conclusions}

With XCT, a new non-destructive method for materials characterization is presented that is complementary to TEM and AFM. TEM investigation requires a destructive sample preparation and can be limited by specimen damage during preparation and measurement, including heat insertion, amorphization, knock-on damage, and radiolysis. Unwanted specimen damage and heating is avoided using XCT, as heat insertion with the method is negligible. XCT offers information on roughnesses of internal and external interfaces, where AFM provides information on external interfaces only. The element specific reflectance contrast in the XUV range also allows a conclusion on the composition without any prior knowledge. By combining a sub-microsecond time scale and a nanometer length scale with XUV contrast based on material composition, the new methodology offers the possibility to study fundamental aspects in materials science, \textit{e.g.}, an \textit{in-situ} measurement of the velocity of off-equilibrium interfaces in the field of rapid phase transformations that at present is by far out of reach for any other approach.

\section*{Acknowledgments}
\noindent This work was supported by the German Research Foundation (grant numbers 390918228 and PA 730/13-1); European Social Fund (ESF) with Thüringer Aufbaubank (2018FGR008, 2015FGR0094), and Bundesministerium für Bildung und Forschung (VIP “X-CoherenT”). J. Apell gratefully acknowledges the State of Thuringia for providing a state scholarship. F. Wiesner is part of the Max Planck School of Photonics supported by BMBF, the Max Planck Society, and the Fraunhofer Society.

\section*{Disclosures}
\noindent The authors declare that there are no conflicts of interest related to this article. M. Wünsche and S. Fuchs are shareholders of the Indigo Optical Systems GmbH.

\bibliographystyle{abbrvnat}
\bibliography{Melting_arxiv}

\end{document}